\def\bold#1{\setbox0=\hbox{$#1$}%
      \kern-.02em\copy0\kern-\wd0
      \kern.04em\copy0\kern-\wd0
      \kern-.02em\raise.0433em\box0 }

\def\bdsmall#1{\setbox0=\hbox{$#1$}%
      \kern-.015em\copy0\kern-\wd0
      \kern.03em\copy0\kern-\wd0
      \kern-.015em\raise.0233em\box0 }

\documentstyle[preprint,amsfonts,prc,aps]{revtex}

\begin{document}
\draft
\title{Helicity asymmetry for proton emission from \\ 
polarized electrons in the eikonal regime}
\author{A. Bianconi}
\address{Dipartimento di Chimica e Fisica per i Materiali, \\
Universit\`a di Brescia, Brescia, Italy I-25133}
\author{M. Radici}
\address{Istituto Nazionale di Fisica Nucleare, Sezione di Pavia, 
Pavia, Italy I-27100}
\date{\today}
\maketitle
\begin{abstract}
The nuclear response to longitudinally polarized electrons, detected in
coincidence with out-of-plane high-energy protons, is discussed in a
simple model where the ejectile wave function is approximated as a plane
wave with a complex wave vector. This choice is equivalent to solve the
problem of Final-State Interactions (FSI) in homogeneous nuclear 
matter, as the residual nucleus can be described to a first 
approximation when dealing with very fast emitted protons. The main 
advantage of the present method is that in the framework of the
Distorted-Wave Impulse Approximation (DWIA) at the one-photon exchange 
level it allows for an analytical derivation of all the components of 
the nuclear response, including the socalled fifth structure function 
$f'_{01}$, which is very sensitive to FSI. The imaginary part of the 
complex wave vector produces purely geometrical FSI effects and, 
consequently, breaks the symmetry of the cross section with respect to 
the incoming electron helicity. Inspection of every single contribution 
in the analytical formulae, here considered up to the fourth order in 
the nonrelativistic reduction in powers of the inverse nucleon mass,  
allows for a detailed study of the role of each elementary reaction 
mechanism. In particular, cancellations among the leading contributions 
determine the very small absolute size of $f'_{01}$ and produce a 
nontrivial asymptotic scaling of the related helicity asymmetry for 
large values of the momentum transfer.
\end{abstract}


\section{Introduction}
\label{sec:intro}

Nuclear reactions induced by electromagnetic probes are well known to
represent a powerful tool to investigate the properties of nuclear 
structure, because the whole target volume can be explored and the
electromagnetic interaction with the external probe is well described 
by the theory of Quantum Electrodynamics 
(QED)~\cite{domb71,bgp85,donn85}. 

In the case of electron scattering, the additional ability of
independently varying energy and momentum transferred to the target, as
well as high-quality beams with large duty factors delivered by modern
electron accelerators, allows for a detailed mapping of the nuclear
response over very different kinematical conditions~\cite{bgprbook}. 

The power of this tool can be better exploited by requiring exclusive
measurements like in the case of $(e,e'p)$ reactions, where the proton 
is detected in coincidence with the final electron and, when possible, 
for a specific energy range corresponding to a well-defined quantum 
state of the residual nucleus. The richness of the structure of the 
theoretical cross section indicates that under suitable kinematical 
conditions it is possible to disentangle observables which are 
selectively sensitive to certain ingredients of the theoretical 
model~\cite{bgprbook,frumou}. 

Additionally, if polarization observables can be measured, it is in
principle possible to determine all the independent scattering 
amplitudes~\cite{simo71}. However, this experimentally formidable goal 
is far from being achieved. More simply, if just the electron beam is 
polarized, it is possible to isolate the socalled fifth structure 
function, which is generated by the interference between two or more 
reaction channels with different competing phases~\cite{bgp85,donn84}. 
Measurements of the corresponding cross section asymmetry with respect 
to electron helicity can be used to explore the interfering reaction 
amplitudes, most of which are usually very small and otherwise hidden 
in the unpolarized case~\cite{oops94}. In electroproduction of pions 
the fifth structure function may provide a key observable for the 
isolation of the resonating channel in the $N \rightarrow \Delta$ 
transition, corresponding to a quadrupole deformed excitation of 
$\Delta^+ (1232)$~\cite{papa89}. Also in inclusive electron scattering 
from polarized targets the helicity asymmetry is needed to access 
observables like the neutron form factor~\cite{donrask86} or the 
spin-dependent nucleon structure functions~\cite{rob90}. 

Here, the completely exclusive quasielastic $(\vec e,e'p)$ reaction on 
nuclear targets will be considered. The fifth structure function is 
then given by the interference between the direct knockout and the 
rescattering channels and is therefore highly sensitive to Final State 
Interactions (FSI) between the outgoing proton and the residual 
nucleus. This issue has become crucial at the new (CEBAF) and planned 
(ELFE~\cite{elfeproj}) high-energy electron accelerators, where 
experiments with electromagnetic probes at momentum transfer beyond 1 
GeV/$c$ (in particular $(e,e'p)$ 
reactions~\cite{cebaf91006,cebaf91007,cebaf91013}) are expected to 
shed some light on exotic phenomena predicted by the perturbative 
Quantum Chromodynamics (pQCD), such as for example the Color 
Transparency (CT). In fact, the experimental signal is predicted to be 
very small in this energy domain and a reliable model for FSI is needed 
to verify the CT prediction~\cite{framilstri94,niko94,bbkha94}. 

In the present literature the most popular and widely adopted approach 
is the Glauber method~\cite{glau59}, which has a long well-established 
tradition of successfull results in the field of high-energy 
proton-nucleus elastic scattering~\cite{pprev78}. Despite the high 
energy regime to which it is applied, this method is developed in a 
completely nonrelativistic formalism within the eikonal approximation. 
Because for a fastly moving object the nuclear density can be 
considered roughly constant inside all the nuclear volume but the small 
part corresponding to the surface, the eikonal wave function of the 
ejectile can be approximated by a damped plane wave, which corresponds 
to the solution of a Schr\"odinger equation inside homogeneous nuclear 
matter. In fact, in a previous paper~\cite{br96a} the angular 
distribution of emitted protons with outgoing energy beyond the 
inelastic threshold and with initially bound momentum below the Fermi 
surface has been shown to be well reproduced by actually assuming a 
plane wave for the final nucleon state with an additional damping. 
Therefore, in the following the scattering wave function will be 
represented as a plane wave with a complex wave vector, whose imaginary 
part produces a constant damping. Consequently, in the framework of 
the Distorted-Wave Impulse Approximation (DWIA) at the one-photon 
exchange level analytical formulae can be derived for all the 
components of the nuclear response. Moreover, the presence of a 
damping in the outgoing plane wave produces a FSI effect of purely 
geometrical nature and generates an asymmetry of the cross section with 
respect to the helicity of the electron beam. 

After a short review on the general formalism (Sec.\  \ref{sec:form}), 
analytical formulae for the fifth structure function and the helicity 
asymmetry will be deduced in Sec.\ \ref{sec:anal}. Results for typical 
kinematics above the inelastic threshold will then be discussed with 
specific emphasys on the asymptotic behaviour for large momentum 
transfer (Sec.\ \ref{sec:out}). Finally, some conclusions will be 
outlined (Sec.\ \ref{sec:end}).

\section{General formalism}
\label{sec:form}

The differential cross section for the scattering of a polarized
electron, with helicity $h$ and initial (final) momentum ${\bold 
p}^{}_{\text{e}} \  ({\bold p}'_{\text{e}})$, off a nuclear
target from which a nucleon is ejected with final momentum 
${\bold p}'$, can be written in the one-photon exchange 
approximation as~\cite{bgprbook} 

\begin{eqnarray}
{ {\text{d}\sigma_h} \over {\text{d}{\bold p}'_{\text{e}} 
\text{d}{\bold p}'} } &= &{ e^4 \over {8 \pi^2}} {1 \over 
{Q^4 p^{}_{\text{e}} p'_{\text{e}}} } \left( 
\rho^{}_{00} f^{}_{00} + \rho^{}_{11} f^{}_{11} + \rho^{}_{01} 
f^{}_{01} \cos \alpha + \rho^{}_{1-1} f^{}_{1-1} \cos 2 \alpha \right. 
\nonumber \\
& &\left. \qquad \qquad \quad \> + h \rho'_{01} f'_{01} \sin 
\alpha \right) \nonumber \\
&\equiv &\Sigma + h \Delta \  , \label{eq:cross}
\end{eqnarray}
where $\alpha$ is the out-of-plane angle (see Fig.\ \ref{fig1}),
$Q^2 = {\bold q}^2 - \omega^2$ and ${\bold q} = 
{\bold p}_{\text{e}} - {\bold p}'_{\text{e}}, \  \omega = 
p^{}_{\text{e}} - p'_{\text{e}}$ are the momentum and energy 
transferred to the target nucleus, respectively. The cross section 
is explicitely separated into a helicity dependent term
$\Delta$ and the helicity independent unpolarized cross section
$\Sigma$. The tensor $\rho_{\lambda \lambda'}$ depends only on 
the properties of the electromagnetic vertex and its 
components are completely determined by QED~\cite{bgp85,donn85}. The 
tensor $f_{\lambda \lambda'}$ contains all the information about the 
target, in particular about the longitudinal $(\lambda=0)$ 
and transverse $(\lambda= \pm 1)$ components of the nuclear 
response with respect to the polarization of the virtual 
photon exchanged. It is given as a bilinear product of matrix 
elements of the different helicity components of the nuclear 
current, which describe the transition from the initial to 
the final hadronic states. Therefore, in principle it involves 
many-body matrix elements. However, in the projection operator 
approach~\cite{bccgp82} and within the framework of DWIA it is 
possible to project out of the total Hilbert space a suitable channel 
where the matrix elements, usually called spectroscopic 
amplitudes, are written in a one-body representation as~\cite{bgprbook} 

\begin{equation}
J^{\lambda}_{nljm_ls's} ({\bold q}) = \displaystyle{\int} \text{d} 
{\bold r} \text{d} \sigma \  \text{e}^{{\scriptstyle \text{i}} 
{\bdsmall {\scriptstyle q}} \cdot {\bdsmall {\scriptstyle r}}} 
\chi^{\left( -\right)\, *}_{s'} ({\bold r}, \sigma) \  
{\hat J}^{}_{\lambda} ({\bold q}, {\bold r}, \sigma) \  
\phi^{}_{nljm_ls} ({\bold r}, \sigma) \> . \label{eq:scattampl}
\end{equation}
They describe the knockout of a nucleon leaving a hole with
quantum numbers $(nljm_ls)$ and propagating across the residual
nucleus with the scattering wave function $\chi^{\left( -
\right)}_{s'}$, $s'$ being the final (detected) spin. The 
normalization of the bound state $\phi$ is the spectroscopic factor, 
which measures the probability that the residual nucleus can 
indeed be considered as a pure hole generated in the target 
nucleus by the knockout process. The boundary conditions for 
the scattering wave $\chi^{\left( - \right)}_{s'}$ are such 
that each incoming partial wave coincides asymptotically with 
the corresponding component of the plane wave associated to 
the outgoing proton momentum ${\bold p}'$ and spin $s'$. 

The amount of $(e,e'p)$ data presently available can be 
explained within the DWIA by adopting for $\phi, 
\chi^{\left( - \right)}$ the solutions of eigenvalue problems
with phenomenological, single-particle, local, energy- and
spin-dependent potentials of the Woods-Saxon type~\cite{bgprbook}. The 
current operator ${\hat J}_{\lambda}$ is usually approximated by a
nonrelativistic expansion in powers of the inverse nucleon
mass by means of the Foldy-Wouthuysen canonical 
transformation~\cite{fw50}. In terms of its controvariant coordinates 
${\hat J}^{\mu} = (\rho, {\bold J})$, it is given up to fourth order 
by~\cite{gp80} 

\begin{eqnarray}
\rho^{\left( 0 \right)} &= &F_1 \> , \qquad  \qquad \qquad  
\rho^{\left( 1 \right)} = 0 \> , \nonumber \\
\rho^{\left( 2 \right)} &= &- {1 \over {8 m^2}} (F_1 + 2 \kappa F_2)
\left( Q^2 + 2 \text{i} {\bold \sigma} \cdot {\bold p} \times {\bold q}
\right) \> , \qquad  \rho^{\left( 3 \right)} = 0 \> , \nonumber \\
\rho^{\left( 4 \right)} &= &{1 \over {16 m^4}} \left( {F_1 \over 24} +
\kappa F_2 \right) \left[ ({\bold p} + {\bold q})^2 + {\bold p}^2 
\right] \left( Q^2 + 2 \text{i} {\bold \sigma} \cdot {\bold p} \times 
{\bold q} \right) + {15 \over {384 m^4}} F_1 \left[ (2 {\bold p} + 
{\bold q} ) \cdot {\bold q} \right]^2 \nonumber \\
& &+ {17 \over {384 m^4}} F_1 \left[ ({\bold p} + {\bold q})^2 + 
{\bold p}^2 \right] \left( q^2 + 2 \text{i} {\bold \sigma} \cdot 
{\bold p} \times {\bold q} \right) \> , \nonumber \\
& & \nonumber \\
{\bold J}^{\left( 0 \right)} &= &0 \> , \qquad \qquad \qquad 
{\bold J}^{\left( 1 \right)} {F_1 \over {2 m}} ( 2 {\bold p} + {\bold
q}) + {{F_1 + \kappa F_2} \over {2 m}} \text{i} {\bold \sigma} \times
{\bold q} \> , \nonumber \\
{\bold J}^{\left( 2 \right)} &= &- {{F_1 + 2 \kappa F_2} \over {8 m^2}}
\text{i} \omega {\bold \sigma} \times ( 2 {\bold p} + {\bold q} ) \> , 
\nonumber \\
{\bold J}^{\left( 3 \right)} &= &- {1 \over {16 m^3}} \left\{ 2 
F_1 \left[ ({\bold p} + {\bold q})^2 + {\bold p}^2 \right] + \kappa 
F_2 Q^2 \right\} - {{\kappa F_2} \over {8 m^3}} \text{i} {\bold \sigma} 
\cdot ( 2 {\bold p} + {\bold q}) {\bold p} \times {\bold q} \> , 
\nonumber \\
{\bold J}^{\left( 4 \right)} &= &{1 \over {16 m^4}} \left( {F_1 \over 
24} + \kappa F_2 \right) \text{i} \omega {\bold \sigma} \times ( 2 
{\bold p} + {\bold q}) \left[ ({\bold p} + {\bold q})^2 + {\bold p}^2 
\right] \nonumber \\
& &+ {17 \over {384 m^4}} F_1 \omega \left[ ({\bold p} + {\bold q})^2 + 
{\bold p}^2 \right] \left[ {\bold q} + \text{i} {\bold \sigma} \times 
(2 {\bold p} + {\bold q}) \right] \nonumber \\
& &+ {15 \over {384 m^4}} F_1 \omega (2 {\bold p} + {\bold q}) \cdot 
{\bold q} \left( 2 {\bold p} + {\bold q} + \text{i} {\bold \sigma} 
\times {\bold q} \right) \> , \label{eq:curr}
\end{eqnarray}
where $m, \kappa$ are the proton mass and anomalous magnetic moment,
respectively, and $F_1, F_2$ are the Dirac, Pauli proton form factors 

\begin{eqnarray}
F_1 (Q^2) &= &\left( 1 + {Q^2 \over {4 m^2}} \right)^{-1} \left[
G_{\text{E}} (Q^2) + {Q^2 \over {4 m^2}} G_{\text{M}} (Q^2) \right] ,
\nonumber \\
\kappa F_2 (Q^2) &= &\left( 1 + {Q^2 \over {4 m^2}} \right)^{-1} \left[
G_{\text{M}} (Q^2) - G_{\text{E}} (Q^2) \right] \label{eq:diracff}
\end{eqnarray}
with $G_{\text{E}}, G_{\text{M}}$ parametrized as in 
Ref.~\cite{bosted95}.  

If the electron beam is polarized, the fifth structure
function $f'_{01}$ enters the cross section as described in
Eq.\ (\ref{eq:cross}) and is given by the following bilinear
product of scattering amplitudes~\cite{bgprbook}

\begin{equation}
f'_{01} = \displaystyle{{{2 q} \over Q}} \ 
{\lower7pt\hbox{$_{s'm_ls\overline{m_l}\overline{s}}$}} 
\kern-24pt {\hbox{\raise2.5pt \hbox{$\sum$}}} \quad   
\left( l \textstyle{1 \over 2} m_l s \vert j m_j \right) 
\left( l \textstyle{1 \over 2} \overline{m_l} \overline{s} 
\vert j m_j \right) \  \text{Im} \{ 
J^0_{nljm_ls's} J^{1\ *}_{nlj\overline{m_l}s'\overline{s}} - 
J^0_{nljm_ls's} J^{-1 \  *}_{nlj\overline{m_l}s'\overline{s}} \} \> .
\label{eq:fifth}
\end{equation}
A necessary condition for having an imaginary component of the
interfering longitudinal-transverse response is the presence
of at least two competing reaction amplitudes with different
phases~\cite{donn84}. As already mentioned, in quasielastic nucleon 
knockout the two dominant channels are the direct emission 
and the rescattering. Therefore, in the absence of any FSI,
the socalled Plane-Wave Impulse Approximation (PWIA), the
$f'_{01}$ identically vanishes because the absence of any
rescattering makes the bilinear products in Eq.\
(\ref{eq:fifth}) purely real and symmetric around ${\bold q}$.
Therefore, the $f'_{01}$ provides a suitable observable to
monitor the rescattering processes in $(\vec e,e'p)$ reactions
and may permit a much higher precision in constraining the
models of FSI by isolating important and otherwise
inaccessible reaction amplitudes. However, this is possible
only at the cost of measuring the reaction products out of the
scattering plane. 

From the experimental point of view it is more advantageous to
isolate the helicity dependent term in Eq.\ (\ref{eq:cross}), 
$\Delta$, which is proportional to $f'_{01}$, by measuring the
asymmetry 

\begin{equation}
A = \  {{\text{d}\sigma_+ - \text{d}\sigma_-} \over 
{\text{d}\sigma_+ + \text{d}\sigma_-}} \  = {\Delta \over 
\Sigma} \> , \label{eq:asymm}
\end{equation}
because the systematic uncertainties in all the
spectrometer efficiencies, target thickness, charge
collection, cancel in the ratio~\cite{oops94}. FSI can break the 
symmetry when flipping the electron helicity $h$ or, equivalently, 
when reaction products are scattered above or below the scattering 
plane for a given $h$. The different path followed on the way out of 
the nucleus makes outgoing protons have different rescatterings with 
the residual and, consequently, produces a phase difference with
respect to the channel where they are knocked out directly as free 
particles. Therefore, at variance with what happens in the case of 
$(\vec e,e')$ where the asymmetry arises from a parity violating 
interaction, here $A$ can be generated for simple geometrical arguments 
just by the modification of the plane wave of the final protons.

\section{A simple model: analytical formulae for 
$\lowercase{f}_{\lambda \lambda'}$}
\label{sec:anal}

For sake of simplicity, we will consider proton knockout from the 
$s\textstyle{1 \over 2}$ shell. No effects will then be produced by the
spin-orbit interaction in the final state, but rather from FSI based on
the simple geometrical arguments mentioned in the previous Section.

In PWIA the scattering amplitude for the knockout from the 
$s\textstyle{1 \over 2}$ shell in configuration space reads 

\begin{equation}
J^{\lambda}_{00\scriptstyle{1 \over 2}0s's} ({\bold q}) = { 
{\lower7pt\hbox{$_{\tilde{s}}$}} \kern-7pt {\hbox{\raise2.5pt
\hbox{$\sum$}}} } \displaystyle{\int} \text{d} {\bold r} \text{d} 
\sigma \  \text{e}^{{\scriptstyle \text{i}} {\bdsmall {\scriptstyle q}} 
\cdot {\bdsmall {\scriptstyle r}}} 
\text{e}^{{\scriptstyle - \text{i}} {\bdsmall {\scriptstyle p'}} 
\cdot {\bdsmall {\scriptstyle r}}} \  \delta^{}_{s'\tilde{s}} \  
\langle \tilde{s} \, \vert  \  {\hat J}^{}_{\lambda} ({\bold q}, 
{\bold r}, \sigma) \  \vert \, s \rangle \  R^{}_{00\scriptstyle{1 
\over 2}} (r) Y^{}_{00} (\Omega_{\bdsmall {\scriptstyle r}})  
\label{eq:scattamplpw}
\end{equation}
and in momentum space it becomes

\begin{eqnarray}
J^{\lambda}_{00\scriptstyle{1 \over 2}0s's} ({\bold q}) &=  &{ 
{\lower7pt\hbox{$_{\tilde{s}}$}} \kern-7pt {\hbox{\raise2.5pt
\hbox{$\sum$}}} } \int \text{d} {\bold p} \text{d} \sigma \  \delta 
({\bold p'} - {\bold p} - {\bold q}) \  \delta^{}_{s'\tilde{s}} \  
\langle \tilde{s} \, \vert  \  {\hat J}^{}_{\lambda} ({\bold q}, 
{\bold p}, \sigma) \  \vert \, s \rangle \  R^{}_{00\scriptstyle{1 
\over 2}} (p) Y^{}_{00} (\Omega_{\bdsmall {\scriptstyle p}}) 
\nonumber \\
&=  &\langle s' \, \vert  \  {\hat J}^{}_{\lambda} ({\bold q}, 
{\bold p'} - {\bold q}, \sigma) \  \vert \, s \rangle \  {1 \over 
{\sqrt{4 \pi}}} R^{}_{00\scriptstyle{1 \over 2}} \left( \vert 
{\bold p'} - {\bold q} \vert \right) \> , \label{eq:scattamplpw1}
\end{eqnarray}
where ${\hat J}^{}_{\lambda} ({\bold q}, {\bold p}, \sigma)$ and 
$R^{}_{00\scriptstyle{1 \over 2}} (p)$ are the Fourier transforms of 
the representation in configuration space of the corresponding 
current operator and radial bound state in Eq.\ (\ref{eq:scattamplpw}), 
respectively. 

The eikonal approximation of the scattering state at the lowest order 
can be represented by a plane wave with a complex momentum 
${\bold P}' = {\bold p'} + \text{i} {\bold p''}$:

\begin{equation}
\text{e}^{- {\bdsmall {\scriptstyle p''}} \cdot {\bdsmall {\scriptstyle
R}}} \  \text{e}^{{\scriptstyle \text{i}} 
{\bdsmall {\scriptstyle P'}} \cdot {\bdsmall {\scriptstyle r}}} = 
\text{e}^{- {\bdsmall {\scriptstyle p''}} \cdot {\bdsmall {\scriptstyle
R}}} \  \text{e}^{{\scriptstyle \text{i}} {\bdsmall 
{\scriptstyle p'}} \cdot {\bdsmall {\scriptstyle r}}} \  
\text{e}^{- {\bdsmall {\scriptstyle p''}} \cdot 
{\bdsmall {\scriptstyle r}}} \> , \label{eq:delta}
\end{equation}
where ${\bold R}$ is a constant vector with modulus equal to the 
nuclear radius. The factor $\text{e}^{- {\bdsmall {\scriptstyle p''}} 
\cdot {\bdsmall {\scriptstyle R}}}$ represents the proper 
normalization. In fact, if $\hat z$ is the propagation axis, the wave 
enters the nucleus at ${\bold r} = - {\bold R} \equiv (0,0,-R)$ with 
unitary modulus and leaves it at ${\bold r} = {\bold R} \equiv (0,0,R)$ 
with the damping $\text{e}^{- 2 {\bdsmall {\scriptstyle p''}} \cdot 
{\bdsmall {\scriptstyle R}}}$. 

By analytically extending the integrand of Eq.\ (\ref{eq:scattamplpw1}) 
into the complex plane ${\bold P}$, it is possible to go beyond the 
PWIA and, at the same time, to perform the integration still 
analytically. The extension to the complex plane has two requirements. 
Firstly, a new definition of the distribution $\delta$ of a complex 
variable (see Appendix), which automatically connects it to the 
``plane'' wave $\text{e}^{{\scriptstyle \text{i}} {\bdsmall 
{\scriptstyle P'}} \cdot {\bdsmall {\scriptstyle r}}}$ of 
Eq.\ (\ref{eq:delta}) in the same way as for the case of a real 
momentum. Secondly, the functions ${\hat J}^{}_{\lambda} ({\bold q}, 
{\bold P}, \sigma), R^{}_{00\scriptstyle{1 \over 2}} (P)$ must be well 
behaved and their product must asymptotically vanish for $P 
\rightarrow \infty$. With these restrictions 
Eq.\ (\ref{eq:scattamplpw1}) can be extended into the complex plane, 
i.e.

\begin{eqnarray}
J^{\lambda}_{00\scriptstyle{1 \over 2}0s's} ({\bold q}) &= \  &{ 
{\lower7pt\hbox{$_{\tilde{s}}$}} \kern-7pt {\hbox{\raise2.5pt
\hbox{$\sum$}}} } \displaystyle{\int} \text{d} {\bold P} \text{d} 
\sigma \  \delta ({\bold P'} - {\bold P} - {\bold q}) \  
\text{e}^{{\scriptstyle \text{i}} \left( {\bdsmall {\scriptstyle P}} + 
{\bdsmall {\scriptstyle q}} - {\bdsmall {\scriptstyle p'}} \right) 
\cdot {\bdsmall {\scriptstyle R}}} \  \delta^{}_{s'\tilde{s}} \  
\langle \tilde{s} \, \vert  \  {\hat J}^{}_{\lambda} ({\bold q}, 
{\bold P}, \sigma) \  \vert \, s \rangle \displaystyle{{1 \over 
{\sqrt{4 \pi}}}} R^{}_{00\scriptstyle{1 \over 2}} (P) \nonumber \\
&= &\langle s' \, \vert  \  {\hat J}^{}_{\lambda} ({\bold q}, 
{\bold P'} - {\bold q}, \sigma) \  \vert \, s \rangle \  
\displaystyle{{1 \over {\sqrt{4 \pi}}}} 
R^{}_{00\scriptstyle{1 \over 2}} \left( \vert {\bold P'} - {\bold q} 
\vert \right) \> , \label{eq:scattampleik}
\end{eqnarray}
where the normalization factor $\text{e}^{- {\bdsmall 
{\scriptstyle p''}} \cdot {\bdsmall {\scriptstyle R}}}$ has been 
included in a redefinition of the bound state 
$R^{}_{00\scriptstyle{1 \over 2}}$.

The scattering wave of Eq.\ (\ref{eq:delta}) represents a simple plane
wave damped by an exponential factor driven by $\text{Im} ({\bold P'}) = 
{\bold p''}$. This corresponds to solve the Schr\"odinger equation with 
a complex potential for a particle travelling through homogeneous 
nuclear matter, i.e. 

\begin{equation}
\left( {- \hbar^2 \over 2m} \nabla^2 + \hat{V} + \text{i} \hat{W} 
\right) \chi = E \chi \> , \label{eq:schro}
\end{equation}
or, equivalently,

\begin{equation}
\left( E - {\hat V} - \text{i} {\hat W} \right) \chi = 
\displaystyle{{ {{\hat {\bold P'}} \cdot {\hat {\bold P'}} } \over 2m}} 
\  = \  \left( \displaystyle{{{{\hat {\bold p}}'^2 - 
{\hat {\bold p}}''^2} \over 2m}} + \text{i} \  \displaystyle{{{{\hat 
{\bold p}}' \cdot {\hat {\bold p}}''} \over m}} \right) \chi \> , 
\label{eq:schroop}
\end{equation}
from which a natural relationship between $p''$ and the absorbitive 
part $W$ of the potential is deduced. If the outgoing proton is 
sufficiently energetic, i.e. $p' \gtrsim 1$ GeV/$c$, and comes from a 
bound state with a momentum below the Fermi surface, this approximation 
has been shown to give reliable results~\cite{br95,br96a,br96b} with a 
constant $p'' \propto W/p'$. Therefore, the question is whether the 
description of FSI by a simple plane wave with a constant damping is 
sufficient to generate an asymmetry in the cross section with respect 
to the incoming electron helicity, i.e. a nonvanishing fifth structure 
function. The answer is positive and analytical formulae will be given 
in the following. 

If the damping of the plane wave is constant not only in size, but
also in its direction, i.e. ${\bold p''} \parallel {\bold p'}$, then

\begin{eqnarray}
{\bold P'} &= &{\bold p'} + \text{i} {\bold p''} = {\bold p'} + 
\text{i} {{p''} \over {p'}} {\bold p'}  =  {\bold p'} \left( 1 + 
\text{i} {{p''} \over {p'}} \right) \nonumber \\
&\equiv &({\bold p}_m + {\bold q}) \left(1 + \text{i} {p'' \over p'} 
\right) \> , \label{eq:mom}
\end{eqnarray}
where ${\bold p}_m$ is the missing momentum of the reaction. By
substituting Eqs.\ (\ref{eq:scattampleik}), (\ref{eq:mom}) in Eq.\
(\ref{eq:fifth}) for the $s\textstyle{1 \over 2}$ knockout shell, the
analytical expression for $f'_{01}$ becomes

\begin{eqnarray}
f'_{01} &= &- {{\sqrt{2} q} \over {\pi Q}} \  R_{00\scriptstyle{1 \over
2}} \left( \vert {\bold p}_m + \text{i} {\bold p''} \vert \right) 
p_{m_{x}} {p'' \over p'} \nonumber \\
& &\left[ S ( \omega, q^2 ) + D ( \omega, q^2, p'^2 , p''^2 ) + D' 
( \omega, q ) ( p_{m_z} + q ) \right] \> , \label{eq:fifthanal}
\end{eqnarray}
where

\begin{eqnarray}
D ( \omega, q^2, p'^2 , p''^2 ) &= &\displaystyle{{{3 \kappa F_1 F_2} 
\over {8 m^3}}} Q^2 + \displaystyle{{{F_1^2} \over m}} \left[ 
\displaystyle{{1 \over {4 m^2}}} \left( 2 p'^2 + 2 p''^2 + q^2 + 
\displaystyle{{1 \over 2}} Q^2 \right) + 
\displaystyle{{5 \over {64 m^3}}} \omega q^2 - 1 \right] \> , 
\nonumber \\
D' ( \omega, q ) &= &- \displaystyle{{{F_1^2} \over {2 m^3}}} q \left( 
1 + \displaystyle{{5 \over {8 m}}} \omega \right) \label{eq:nospfl} 
\end{eqnarray}
come from the part of the current operator ${\hat J}^{}_{\lambda} 
({\bold q}, {\bold P'} - {\bold q}, \sigma)$ which does not flip the
initial nucleon spin, while 

\begin{equation}
S ( \omega, q^2 ) = {{F_1 + 2 \kappa F_2} \over {8 m^3}} q^2 \left[ 
F_1 + \kappa F_2 + {\omega \over {4 m}} \left( F_1 + 2 \kappa F_2 
\right) \right] \label{eq:spfl}
\end{equation}
is produced by the spin-flip part. Here, the vector components along 
$\hat x$, $\hat z$ are referred to the hadronic plane $({\bold p'}, 
{\bold q})$. If $\alpha=0^{\text{o}}$ the latter coincides with the 
scattering plane and $p_{m_x}$ actually represents the component along 
the ${\hat x}$ axis of the lab system described in Fig.\ \ref{fig1}. 
If $\alpha=90^{\text{o}}$ the hadronic plane is perpendicular to the 
scattering plane and $p_{m_x}$ refers to the component along the 
${\hat y}$ axis of the lab system. No ambiguity should arise from the 
interpretation of the components along the ${\hat z}$ axis which always 
points in the direction of ${\bold q}$.

If the experimental setup is such that the spectrometer of the hadron 
arm detects the outgoing protons on a plane perpendicular to the 
scattering plane, i.e. for $\alpha = 90^{\text{o}}$ in Fig.\ 
\ref{fig1}, from Eqs.\ (\ref{eq:cross}), (\ref{eq:asymm}) the helicity 
asymmetry takes the simple form

\begin{equation}
A = { {\rho'_{01} f'_{01}} \over {\rho^{}_{00} f^{}_{00} + \rho^{}_{11} 
f^{}_{11} - \rho^{}_{1-1} f^{}_{1-1}} } \quad {\lower5pt\hbox{$_{Q \to 
\infty}$}} \kern-20pt {\hbox{\raise2.5pt \hbox{$\longrightarrow$}}} 
\quad { {\rho'_{01} f'_{01}} \over {\rho^{}_{11} f^{}_{11}} } \> , 
\label{eq:asymm1}
\end{equation}
because for increasing $Q$ the nuclear response becomes more and more
transverse. Since, analogously to Eq.\ (\ref{eq:fifth}), the structure 
function $f^{}_{11}$ is given in terms of the scattering amplitudes
as~\cite{bgprbook} 

\begin{equation}
f^{}_{11} = {\lower7pt\hbox{$_{s'm_ls\overline{m_l}\overline{s}}$}} 
\kern-24pt {\hbox{\raise2.5pt \hbox{$\sum$}}} \quad \  \left( l 
\textstyle{1 \over 2} m_l s \vert j m_j \right) \left( l \textstyle{1 
\over 2} \overline{m_l} \overline{s} \vert j m_j \right) \  
\{ J^1_{nljm_ls's} J^{1\ *}_{nlj\overline{m_l}s'\overline{s}} + 
J^{-1}_{nljm_ls's} J^{-1 \  *}_{nlj\overline{m_l}s'\overline{s}} \} 
\label{eq:second}
\end{equation}
and the components of the lepton tensor read~\cite{bgprbook}

\begin{equation}
\rho'_{01} = {Q^2 \over q^2} {1 \over \sqrt{2}} \tan 
\displaystyle{\theta \over 2} \> , \qquad \qquad 
\rho^{}_{11} = {Q^2 \over {2 q^2}} + \tan^2 
\displaystyle{\theta \over 2} \> , \label{eq:lepton}
\end{equation}
the final analytical expression of the helicity asymmetry for the 
knockout from the $s\textstyle{1 \over 2}$ shell is 

\begin{eqnarray}
A &= &- { {4 q Q \tan \textstyle{\theta \over 2}} \over {Q^2 + 2 q^2 
\tan^2 \textstyle{\theta \over 2}} } \  p_{m_x} {p'' \over p'} \  
\left[ S ( \omega, q^2 ) + D ( \omega, q^2, p'^2 , p''^2 ) + 
D' ( \omega, q ) ( p_{m_z} + q ) \right] \nonumber \\
& &\times \left[ \overline{S} ( \omega, q^2, p'^2, p''^2 ) + 
\overline{S}' ( \omega, q ) ( p_{m_z} + q ) + \overline{S}'' 
( \omega^2, p'^2, p''^2 ) ( p_{m_z} + q )^2  \right. \nonumber \\
& &\left. \quad \  + \overline{D} ( \omega^2, q^2, p'^2, p''^2 ) 
( p^2_{m_x} + p^2_{m_y} ) + \overline{D}' ( q, p'^2, p''^2 ) 
( p_{m_z} + q ) ( p^2_{m_x} + p^2_{m_y} ) \right]^{-1} \> . 
\label{eq:asymmanal}
\end{eqnarray}
Analogously to Eqs.\ (\ref{eq:nospfl}), (\ref{eq:spfl}) the functions
$\overline{S}, \overline{S}', \overline{S}''$ and $\overline{D}, 
\overline{D}'$ are produced by the spin flipping and non spin flipping 
parts of the interaction, respectively. Their expressions are

\begin{eqnarray}
&\overline{S} ( \omega, q^2, p'^2, p''^2 ) = \left( \displaystyle{{ 
{F_1 + 2 \kappa F_2} \over {4 m^2}}} \omega \right)^2 \left( p'^2 + 
p''^2 \right) + \displaystyle{{q^2 \over {2 m^2}}} \left[ F_1 + 
\kappa F_2 + \displaystyle{{ {F_1 + 2 \kappa F_2} \over {4 m} }} 
\omega \right]^2 \qquad \nonumber \\
&\qquad \qquad \quad - \displaystyle{{ {F_1 + \kappa F_2} \over {8 m^4} 
}} q^2 \left[ 4 F_1 (p'^2 + p''^2) + 2 F_1 q^2 + \kappa F_2 Q^2 \right] 
\> , \quad \qquad \qquad \nonumber \\
&\overline{S}' ( \omega, q ) = \displaystyle{{q \over {2 m^3}}} \left[ 
\displaystyle{{F_1 \over m}} (F_1 + \kappa F_2) q^2 - (F_1 + \kappa 
F_2) (F_1 + 2 \kappa F_2) \omega - \displaystyle{{ {(F_1 + 2 \kappa 
F_2)^2} \over {4 m}}} \omega^2 \right] \> , \nonumber \\
&\overline{S}'' ( \omega^2, p'^2, p''^2 ) = \left( 
\displaystyle{{{F_1 + 2 \kappa F_2} \over {4 m^2}}} \omega \right)^2 
\left( 1 + \displaystyle{{p''^2 \over p'^2}} \right) \> , 
\qquad \qquad \qquad \qquad \qquad \qquad \qquad \quad
\nonumber \\
&\overline{D} ( \omega^2, q^2, p'^2, p''^2 ) = \displaystyle{{F_1 
\over {m^2}}} \left( 1 + \displaystyle{{p''^2 \over p'^2}} \right) 
\left[ F_1 - \displaystyle{{F_1 \over {m^2}}} (p'^2 + p''^2)
-  \displaystyle{{F_1 \over {2 m^2}}} q^2 - \displaystyle{{{\kappa F_2} 
\over {4 m^2}}} Q^2 \right] \> , \quad \  \  \nonumber \\
&\overline{D}' ( q, p'^2, p''^2 ) = \displaystyle{{F^2_1 \over m^4}} q 
\left( 1 + \displaystyle{{p''^2 \over p'^2}} \right) \> .
\qquad \qquad \qquad \qquad \qquad \qquad \qquad \qquad \qquad
\qquad \quad \label{eq:denom} 
\end{eqnarray}

Finally, it should be noticed, from Eqs.\ (\ref{eq:fifthanal}), 
(\ref{eq:asymmanal}) respectively, that both $f'_{01}$ and $A$ depend 
on $p''/p'$, which is related to the imaginary part of the complex 
momentum ${\bold P}'$ defined in Eq.\ (\ref{eq:mom}). This ratio gives 
a measure of the damping of the scattering wave, i.e. of the FSI. In 
fact, for $p'' \rightarrow 0$ the damping vanishes: the scattering 
wave becomes a plane wave with momentum ${\bold P}' \equiv {\bold p}'$ 
and $f'_{01}, A$ vanish, as it must be in PWIA.

\section{Results}
\label{sec:out}

In this Section results will be shown for the fifth structure function 
$f'_{01}$ of Eq.\ (\ref{eq:fifthanal}) and for the helicity asymmetry 
$A$ of Eq.\  (\ref{eq:asymmanal}) for the proton knockout from a 
$s\textstyle{1 \over 2}$ shell by a polarized electron beam. For sake 
of consistency with previous calculations~\cite{br95,br96a,br96b} and 
available measurements~\cite{oops94,ne18}, the $^{12}\text{C}$ target 
has been selected. The choice of the residual $^{11}\text{B}$ with 
quantum numbers $s\textstyle{1 \over 2}$ is justified, as mentioned in 
the previous Section, by the absence of any FSI due to spin-orbit 
effects, which not only makes formulae simpler, but also clarifies the 
pure geometrical nature of FSI introduced by a plane wave with complex 
wave vector. Because of the purely absorbitive nature of the damping, 
the energy range available to the final proton has been selected 
above the inelastic threshold of $p' \sim 1$ GeV/$c$. 

It has been shown elsewhere~\cite{br96b} that the nuclear response for 
$p^{}_m$ well above the target Fermi momentum $p^{}_{\text{Fermi}}$ is 
dominated by FSI with a nontrivial structure, while for $p^{}_m 
\lesssim p^{}_{\text{Fermi}}$ it can be described as the PWIA 
contribution with an additional constant damping. Since the 
propagation of the outgoing proton with a complex wave vector can 
actually be pictured as a plane wave with a constant damping, it seems 
natural to select values of $p^{}_m$ inside a range where the adopted 
representation of FSI is not too inadequate. The Fermi momentum of 
$^{12}\text{C}$ is $p^{}_{\text{Fermi}} = 221$ MeV/$c$; therefore, in 
the following, a typical value of $p^{}_m = 200$ MeV/$c$ will be
used. 

In the considered domain of inelastic processes $p' \sim q \gg p^{}_m$. 
Therefore, the kinematics is almost purely transverse. In the 
following, without loss in generality, it will be kept exactly 
transverse, i.e. with $p_{m_z} = 0, p_{m_x} = 200$ MeV/$c$. The 
damping factor $p''$ has been shown, in the previous Section, to be 
directly related to the the imaginary part $W$ of the equivalent 
optical potential. If the Glauber choice of $W \propto p'$ is adopted, 
the observed damping in the NE18 data is reproduced by selecting $V=0, 
W = 0.036 \  p'$ MeV~\cite{br96a,br96b}. Correspondingly, in the 
following $p'' = 50$ MeV/$c$ will be used. 

As a cross-check, the helicity asymmetry of Eq.\ (\ref{eq:asymmanal}) 
for the present choices of ${\bold p}_m$ and $p''$ has been compared 
in the range $0.6 \leq q \leq 1$ GeV/$c$ and for a quasielastic 
kinematics ($\omega \simeq q^2/2 m$) with the output of the numerical 
code {\tt PV5FF} developed in Pavia, which successfully describes the 
amount of 
presently available $(e,e'p)$ and $(\vec e, e'p)$ data at medium proton 
energies in the framework of DWIA and including spin-dependent FSI and 
Coulomb distortion of the electron waves~\cite{bgprbook}. In Fig.\ 
\ref{fig2} the solid line corresponds to the analytical formula, while 
the dashed line is the numerical result obtained with the complex 
optical potential $V=0, W= 0.036 \  p'$ MeV and the bound state of 
Comfort and Karp~\cite{comf80} for the $s\textstyle{1 \over 2}$ shell 
of $^{12}\text{C}$. The agreement is satisfactory for $q \lesssim 0.8$ 
GeV/$c$, while above this threshold the accuracy required by the 
delicate cancellations taking place in the numerator of 
Eq.\ (\ref{eq:asymm1}) is not fulfilled by the numerical code, which 
was optimized for lower energies.

\subsection{The fifth structure function $f'_{01}$}
\label{sec:fifth}

In Eq.\ (\ref{eq:fifthanal}) emphasys has been put on identifying the 
single contributions coming from different reaction mechanisms 
(flipping or non flipping of the nucleon spin) to put in better 
evidence the delicate interplay that leads to a very small structure 
function. 

In Fig.\ \ref{fig3} the $f'_{01}$ (apart from the constant factor 
$- \sqrt{2} R^2_{00\scriptstyle{1 \over 2}} \left( \vert {\bold p}_m + 
\text{i} {\bold p''} \vert \right) / \pi$) is represented by the solid 
line for the 
$^{12}\text{C}(\vec e,e'p)^{11}\text{B}_{s\scriptstyle{1 \over 2}}$ 
reaction as a function of $q$ with $p_{m_x}=200$ MeV/$c$ and $p''=50$ 
MeV/$c$. The results for the functions 
$D ( \omega, q^2, p'^2 , p''^2 ), D' ( \omega, q ), S ( \omega, q^2 )$ 
of Eqs.\ (\ref{eq:nospfl}), (\ref{eq:spfl}) are indicated by the 
short-dashed, long-dashed and dot-dashed lines, respectively. It should 
be noticed that the total result is amplified by a factor $10^2$ with 
respect to each addendum. This dramatic cancellation is the natural 
counterpart of $f'_{01}$ being defined as the difference of 
contributions coming from the interference between longitudinal 
$(\lambda=0)$ and transverse $(\lambda=\pm 1)$ components of the 
nuclear current (see Eq.\  (\ref{eq:fifth})). This peculiar feature on 
one side makes $f'_{01}$ very interesting because extremely sensitive 
to reaction channels emphasized in the interference (to FSI, in this 
case of quasielastic knockout), but on the other side produces a very 
small, hardly measurable quantity.

Experimentally, it is possible to directly determine $f'_{01}$ by 
performing an absolute measurement of the corresponding unpolarized 
cross section $\Sigma$ and of the helicity asymmetry $A$~\cite{oops94}. 
The knowledge of $\Sigma$ and $A$ in Eq.\  (\ref{eq:asymm}) makes it 
possible to isolate the helicity dependent part of the cross section, 
$\Delta$, and consequently the fifth structure function through the 
relation

\begin{equation}
f'_{01} = A \Sigma \  {{8 \pi^2 Q^4 p^{}_{\text{e}} p'_{\text{e}}} 
\over {e^4 \rho'_{01}} } \> . \label{eq:fifthexp}
\end{equation}

\subsection{The helicity asymmetry $A$}
\label{sec:asymm}

In Fig.\ \ref{fig4}a the helicity asymmetry $A$ of Eq.\ 
(\ref{eq:asymmanal}) is plotted as a function of $q$ for the 
$^{12}\text{C}(\vec e,e'p)^{11}\text{B}_{s\scriptstyle{1 \over 2}}$ 
reaction with $p_{m_x}=200$ MeV/$c$, $p''=50$ MeV/$c$ and 
$\theta=40^{\text{o}}$. Again, because of the cancellations occurring 
inside $f'_{01}$ the asymmetry quickly becomes very small. A zoom of 
it is given in Fig.\  \ref{fig4}b, which shows an interesting structure 
with a change of sign and a long asymptotic tail. Despite the fact that 
the asymmetry measurement is an experimentally favourite situation, the 
absolute size of $A$ is probably too small to be ever detected. 

However, it is interesting to study the asymptotic behaviour of this
smooth dependence upon $q$, or equivalently $Q$. It has already been 
mentioned that for increasing $Q$ the response to an electron probe is 
known to become more and more transverse with respect to the helicity 
of the virtual photon exchanged. In pQCD simple dimensional 
arguments~\cite{brolep81} show that for exclusive processes like 
elastic electron-proton scattering the ratio between the Dirac and 
Pauli proton form factors, $F_1 / F_2$, scales as $Q^2$. At the cross 
section level this corresponds to the linear scaling in $1/Q$ of the 
ratio $J^0/J^{\pm 1}$, where $J^0$ $(J^{\pm 1})$ is the helicity 
amplitude for absorption by a proton of a longitudinally (transversely) 
polarized photon. Apart from kinematical factors, the fifth structure 
function $f'_{01}$ is approximately a linear combination of products
$J^0 J^{\pm 1}$, while the dominant purely transverse structure 
function $f_{11}$ is essentially given by $(J^{\pm 1})^2$. Therefore, 
one would naively deduce from Eq.\ (\ref{eq:asymm1}) that the helicity 
asymmetry itself asymptotically scales as $1/Q$. 

But the $f'_{01}$ is not just a linear combination of products 
$J^0 J^{\pm 1}$, as it is evident from Eq.\ (\ref{eq:fifth}). The
cancellations between contributions of the same order in powers of 
$1/Q$ are very sensitive to the relativistic corrections to the 
current operator and produce a nontrivial scaling law. Assuming that 
for large $Q^2$ the Bjorken variable $x = Q^2/2m\omega$ is 
approximately constant and, consequently, $\omega \sim q \sim Q^2$, 
from Eqs.\ (\ref{eq:diracff}), (\ref{eq:nospfl}), (\ref{eq:spfl}) and 
(\ref{eq:denom}) it can be deduced that the helicity asymmetry of 
Eq.\ (\ref{eq:asymmanal}) scales as 

\begin{equation}
A \  {\lower5pt\hbox{$_{Q \to \infty}$}} \kern-16pt 
{\hbox{\raise2.5pt \hbox{$\sim$}}} \quad  {1 \over Q^5} \> . 
\label{eq:scaling}
\end{equation}

In the energy domain pertinent to the planned ELFE 
setup~\cite{elfeproj}, the previous assumptions on 
$x, \  q/Q^2, \  \omega/Q^2$ do not hold yet. A different tail as 
power of $1/Q$ must be expected for $A$. In fact, in Fig.\ \ref{fig5} 
the helicity asymmetry is shown, multiplied by $Q^4$, as a function of 
$q$ for the same reaction and in the same kinematical conditions as 
in the previous figure. The plateau indicates that in this energy 
window the scaling is different from what is predicted by 
Eq.\  (\ref{eq:scaling}), or, in other words, that the asymptotic 
behaviour is not yet reached within the present nonrelativistic 
reduction of the current operator at the order described in 
Eq.\ (\ref{eq:curr}) (see also Ref.~\cite{bb95}). 

Finally, since the kinematics is here purely transverse and for high 
$p', q$ the longitudinal component of the missing momentum, $p_{m_z}$, 
is anyway small, the asymmetry is approximately linearly dependent on 
$p_{m_x}$ ($p_{m_y}=0$) and, consequently, does not show any 
interesting structure with respect to $p_{m_z}$.

\section{Conclusions}
\label{sec:end}

The 
$^{12}\text{C}(\vec e ,e'p)^{11}\text{B}_{s\scriptstyle{1 \over 2}}$ 
reaction has been analyzed assuming for the scattering state a plane 
wave with complex wave vector. This choice allows for obtaining 
analytical formulae for the different components of the nuclear 
response; it corresponds to the situation where the outgoing proton 
emerges as a free particle but its wave function is exponentially 
damped with a rate related to the imaginary part $p''$ of the complex 
wave vector. This picture is also equivalent to solve the problem for
the scattering state in the lowest-order eikonal approximation or,
alternatively, to compute the FSI of the outgoing proton travelling 
across absorbitive homogeneous nuclear matter represented by a complex 
potential. In fact, $p''$ has been shown to be directly related to the 
imaginary part of this potential and is of the same order of magnitude. 

Since the residual nucleus is left with quantum numbers $s\textstyle{1
\over 2}$, there are no FSI due to spin-orbit effects. The modification
of the emerging plane wave is the only reason why the symmetry between
protons emitted above and below the scattering plane is broken. The
different path followed on the way out of the nucleus makes them have 
different rescatterings with the residual and, consequently, produces a 
phase difference with respect to the channel where they are knocked out 
directly as free particles. In these conditions, and in general 
whenever there are at least two predominant reaction channels with 
different phases~\cite{donn84}, the cross section part depending on 
the electron helicity $h$ does not vanish.

In particular, the fifth structure function $f'_{01}$ can be used to
disentangle interfering processes and, in the present case of
quasielastic kinematics, to study the rescattering amplitudes. The 
$f'_{01}$ has been analyzed for the previously 
mentioned reaction in purely transverse kinematics for the energy range 
above the inelastic threshold ($p' \sim q \gtrsim 1$ GeV/$c$), where 
the FSI are almost purely absorbitive and the eikonal approximation is 
known to be reliable~\cite{br95,br96a,br96b}. Inspection of the
analytical formula shows that dramatic cancellations take place among 
the different contributions coming from the (non) spin-flipping parts 
of the interaction current. As a result, the absolute size of $f'_{01}$ 
is very small and, presumably, hardly observable.

However, the asymmetry between particles detected above and below the
scattering plane is equivalent to the asymmetry for particles emitted 
in the same direction but for opposite $h$. The helicity dependent 
cross section $\Delta$, proportional to $f'_{01}$, can be singled out 
by making coincidence measurements with a fixed spectrometer at an 
angle out of scattering plane and flipping the helicity of the incoming 
polarized electron. High precision data can be obtained with this 
asymmetry technique, because most systematic errors cancel in the 
ratio~\cite{oops94}. 

The analytical formula for the helicity asymmetry $A$ has been studied 
in the same previous kinematics, specifically focussing on its 
asymptotic behaviour for very large energy and momentum transfer. In 
fact, despite of its very small absolute size, it shows an interesting 
structure with a change of sign and a long asymptotic tail. 

The occurrence inside $f'_{01}$ of cancellations between competing 
contributions, asymptotically scaling with the same power of $1/Q$, is 
very sensitive to higher-order relativistic corrections to the current 
operator and produces in the related $A$ a nontrivial scaling law for 
large $Q$, which cannot be naively deduced from dimensional arguments 
applied to the elementary photo-quark reaction 
amplitudes~\cite{brolep81}. Moreover, this asymptotic scaling occurs 
for very large values of $Q$ outside the range available to the 
operational or planned setups of modern electron accelerators, such as 
CEBAF or ELFE. In particular, in the energy domain of the 
latter~\cite{elfeproj} the asymmetry $A$ still shows a scaling 
behaviour, but with a power law in $1/Q$ which is different from the 
asymptotic one. 

In summary, within the present nonrelativistic reduction of the current
operator at the order described in Eq.\ (\ref{eq:curr}), the helicity
asymmetry is very small in its absolute size but shows a long 
nontrivial tail for large $Q$. For $Q \rightarrow \infty$ it scales as 
$1/Q^5$, but it approaches the asymptotic regime very slowly, even 
locally showing, for large but finite $Q$, different scaling 
behaviours.

\acknowledgments

We would like to thank prof. F. Capuzzi for many clarifying discussions 
on the mathematical definitions described in the Appendix. 

This work has been performed in part under the contract ERB 
FMRX-CT-96-0008 within the frame of the Training and Mobility of 
Researchers Programme of the Commission of the European Communities.

\appendix
\section*{}

The usual distribution $\delta$ can be defined as 

\begin{equation}
\delta (x - \overline{x}) = \lim_{\varepsilon \to 0} \  {\varepsilon 
\over {\left( x - \overline{x} \right)^2 + \varepsilon^2}} = 
\lim_{\varepsilon \to 0} \  {1 \over {2 \pi \text{i}}} \  \left\{ 
{1 \over {x - \overline{x} - \text{i} \varepsilon}} -
{1 \over {x - \overline{x} + \text{i} \varepsilon}} \right\} \> , 
\label{eq:deltare}
\end{equation}
where $x,\overline{x} \in \text{I} \kern -2pt \text{\bf R}$. 
This definition can be generalized to the case of the distribution 
$\delta$ of the complex variable $z$ as~\cite{davy}

\begin{equation}
\delta (z - \overline{z}) = \lim_{\varepsilon \to 0} \  {1 \over {2 \pi
\text{i}}} \  \left\{ {1 \over {z - \overline{z} - \text{i} 
\varepsilon}} - {1 \over {z - \overline{z} + \text{i} \varepsilon}} 
\right\} \> . \label{eq:deltacom}
\end{equation}

The new definition of Eq.\ (\ref{eq:deltacom}) keeps the usual 
properties of the $\delta$, in particular 

\begin{equation}
\int_C \  \text{d} z \  \delta (z - \overline{z}) f(z) = 
f(\overline{z}) \> , \label{eq:intdeltf}
\end{equation}
where $C$ is a integration path on the complex plane, extending to
$\text{Re} (z) \rightarrow \pm \infty$ on the real axis but going 
through the point $z = \overline{z}$, and $f(z)$ is an analytical 
complex function with the property $f(z) \rightarrow 0$ for $\vert z 
\vert \rightarrow \infty, \  \text{Im} (z) > 0 \> (\text{Im} (z) < 0)$ 
if $C$ is closed in the upper (lower) part of the complex plane. 

From Eq.\ (\ref{eq:intdeltf}) it follows that 

\begin{equation}
\int_C \  \text{d} z \  \delta (z - \overline{z})
\text{e}^{\scriptstyle{\text{i} x z}} = \text{e}^{\scriptstyle{\text{i} 
x \overline{z}}} \> , \label{eq:intdelteik}
\end{equation}
which generalizes the standard relationship between the $\delta$ and 
the plane wave through the Fourier transformation. Eq.\
(\ref{eq:intdelteik}) can be demonstrated by closing the path $C$ 
with a semicircle in the upper part of the complex
plane $(\vert z \vert \rightarrow \infty, \  \text{Im} (z) > 0)$ for $x 
\geq 0$ or in the lower part $(\vert z \vert \rightarrow \infty, \  
\text{Im} (z) < 0)$ for $x < 0$.

%%%%%%%%%%%%%%%%%%%%%%%%%%%%% bibliography %%%%%%%%%%%%%%%%%%%%%%%

%%%%%%%%%%%%%%%%%%%%%%%%%%%  Figure Captions %%%%%%%%%%%%%%%%%%%%%

\begin{figure}
\caption{The kinematics for the one-nucleon knockout process from a
polarized electron beam.}
\label{fig1}
\end{figure}

\begin{figure}
\caption{
The helicity asymmetry $A$, multiplied by $10^3$, as a 
function of the momentum transfer $q$ in the range $0.6 \leq q \leq 1$ 
GeV/$c$ for the 
$^{12}\protect{\text{C}}(\vec{e},e'p)^{11}\protect{\text{B}}_{s 
\scriptstyle{1 \over 2}}$ reaction in quasielastic completely 
transverse kinematics ($p_m \equiv p_{m_x} = 200$ MeV/$c$) at the 
scattering angle $\theta = 40^{\protect{\text{o}}}$. The solid line is 
the outcome of the analytical formula of 
Eq.\ (\protect{\ref{eq:asymmanal}}) with the damping $p'' = 50$ MeV/$c$ 
(see text). The dashed line is the output of the numerical
code {\tt PV5FF} based on the DWIA with a purely imaginary 
optical potential with depth $W = 0.036 \  p'$ MeV  and with 
the bound state obtained from the potential of Comfort and 
Karp (see text).
}
\label{fig2}
\end{figure}

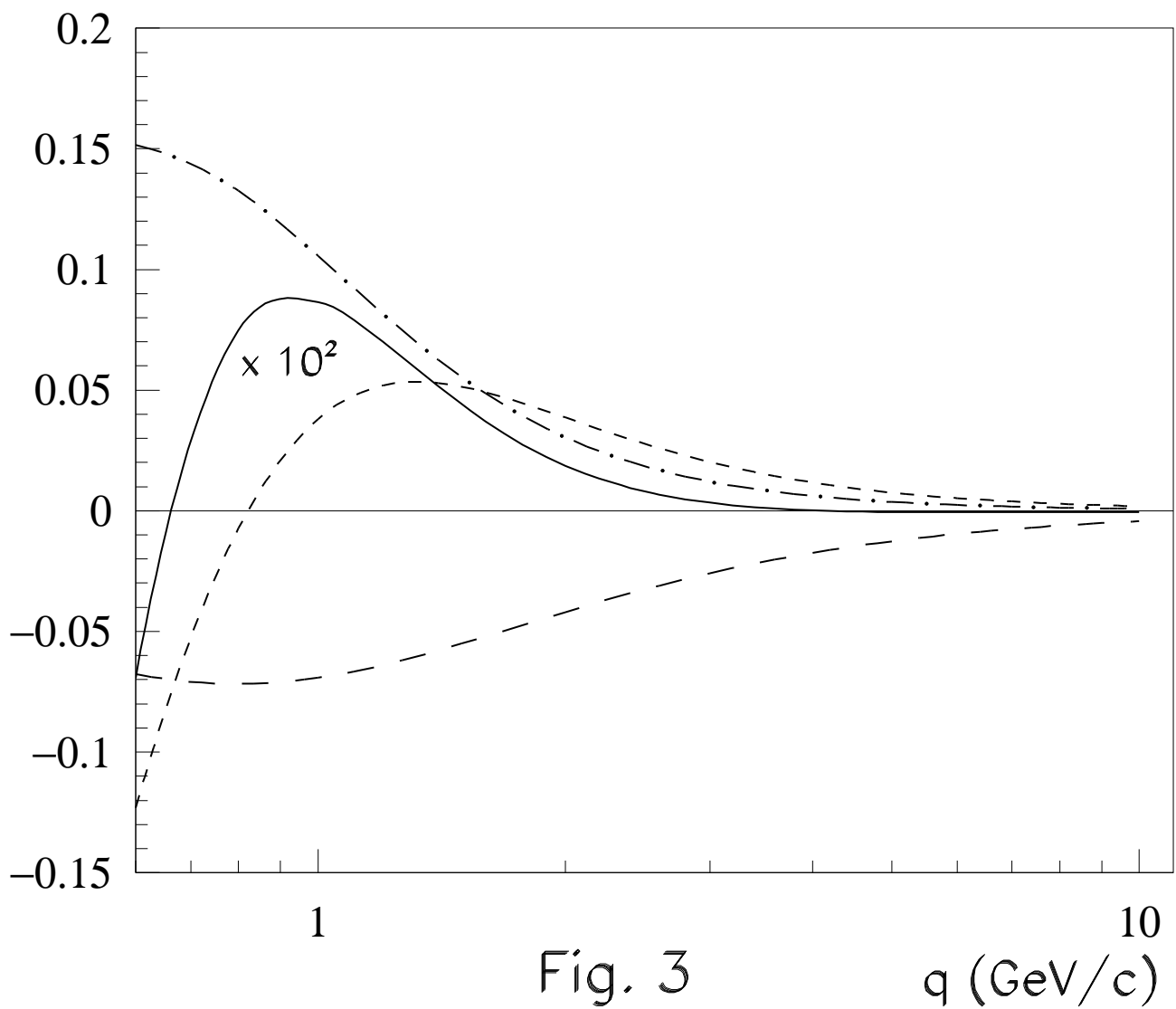
\begin{figure}
\caption{The solid line is the fifth structure function $f'_{01}$ of 
Eq.\  (\protect{\ref{eq:fifthanal}}), multiplied by $10^2$ and divided 
by $- \sqrt{2} R^2_{001/2} /\pi$, as a function of the momentum 
transfer $q$ for the 
$^{12}\protect{\text{C}}(\vec e,e'p)^{11}\protect{\text{B}}_{s 
\scriptstyle{1 \over 2}}$ reaction in the same kinematical conditions 
as for the solid line in Fig.\ \protect{\ref{fig2}}. The short-, long- 
and dot-dashed lines are the $D ( \omega, q^2, p'^2 , p''^2 ), D' 
( \omega, q )$ and $S ( \omega, q^2 )$ functions of Eqs.\ 
(\protect{\ref{eq:nospfl}}), (\protect{\ref{eq:spfl}}), respectively.}
\label{fig3}
\end{figure}

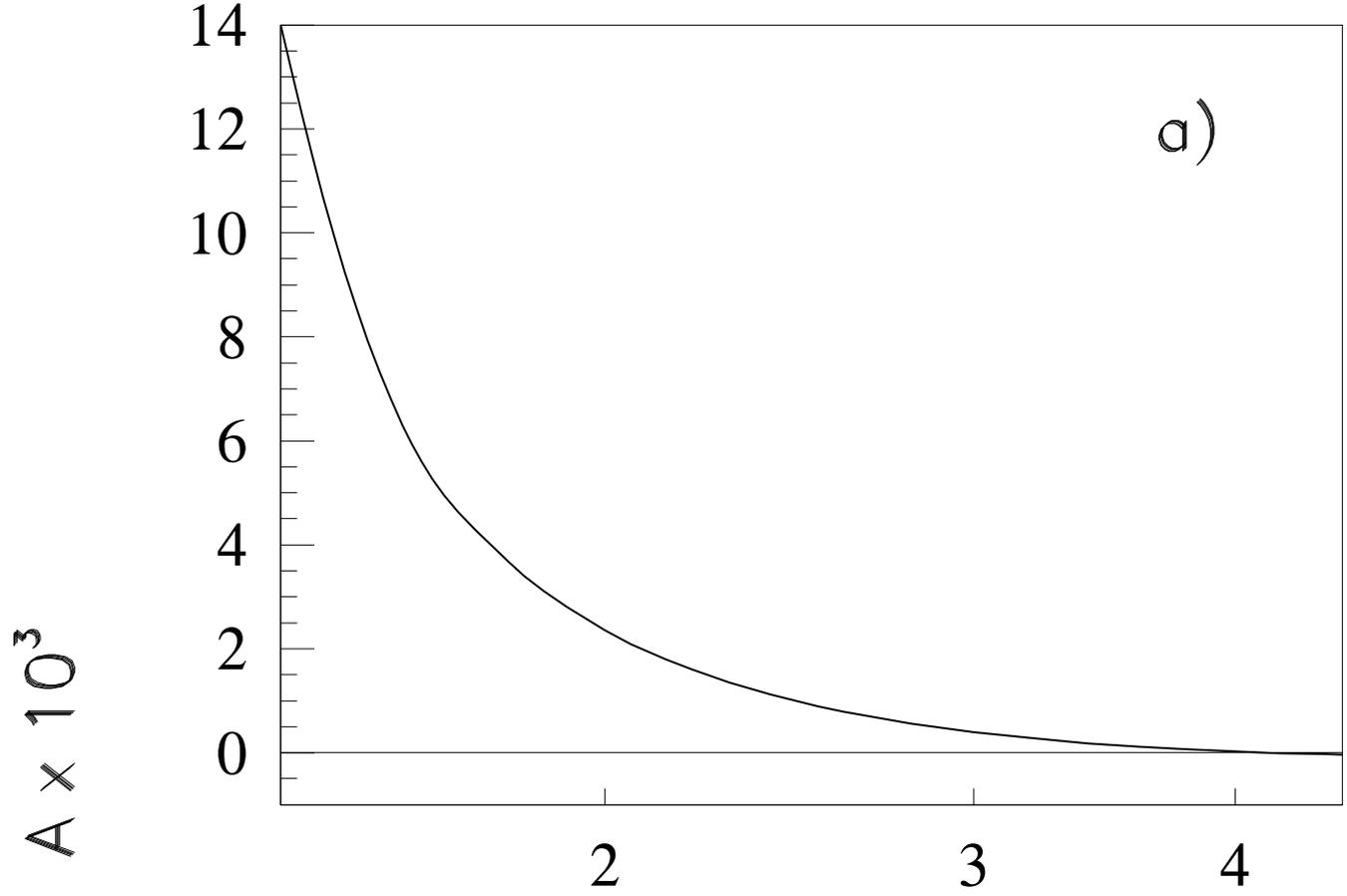
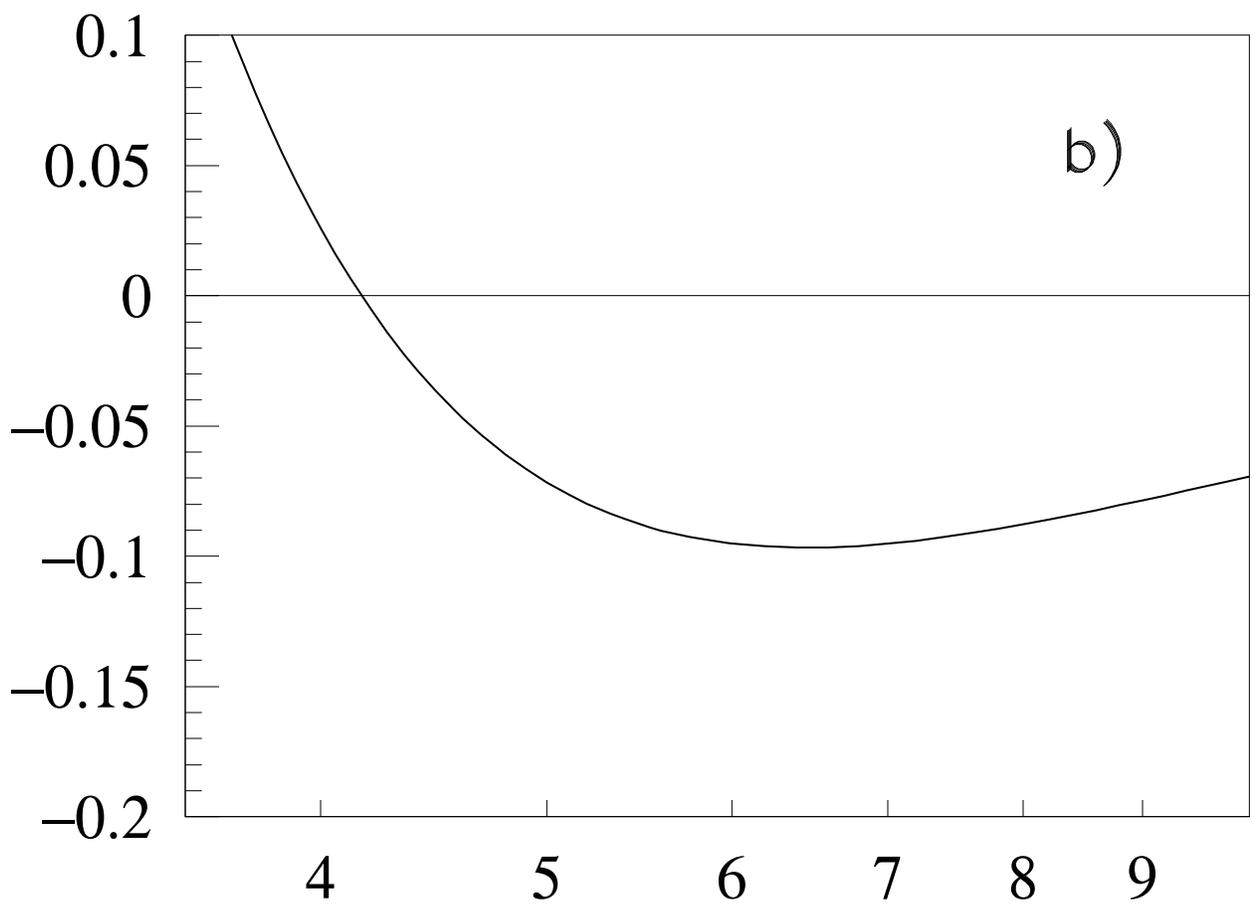
\begin{figure}
\caption{The helicity asymmetry $A$, multiplied by $10^3$, as a 
function of the momentum transfer $q$ for the 
$^{12}\protect{\text{C}}(\vec e,e'p)^{11}\protect{\text{B}}_{s 
\scriptstyle{1 \over 2}}$ reaction in the same kinematical conditions 
as for the solid line in Fig.\ \protect{\ref{fig2}}. Upper part (a) 
for the range $1.4 \leq q \leq 4.5$ GeV/$c$, lower part (b) for the 
range $3.5 \leq q \leq 10$ GeV/$c$ and in an amplified scale.}
\label{fig4}
\end{figure}

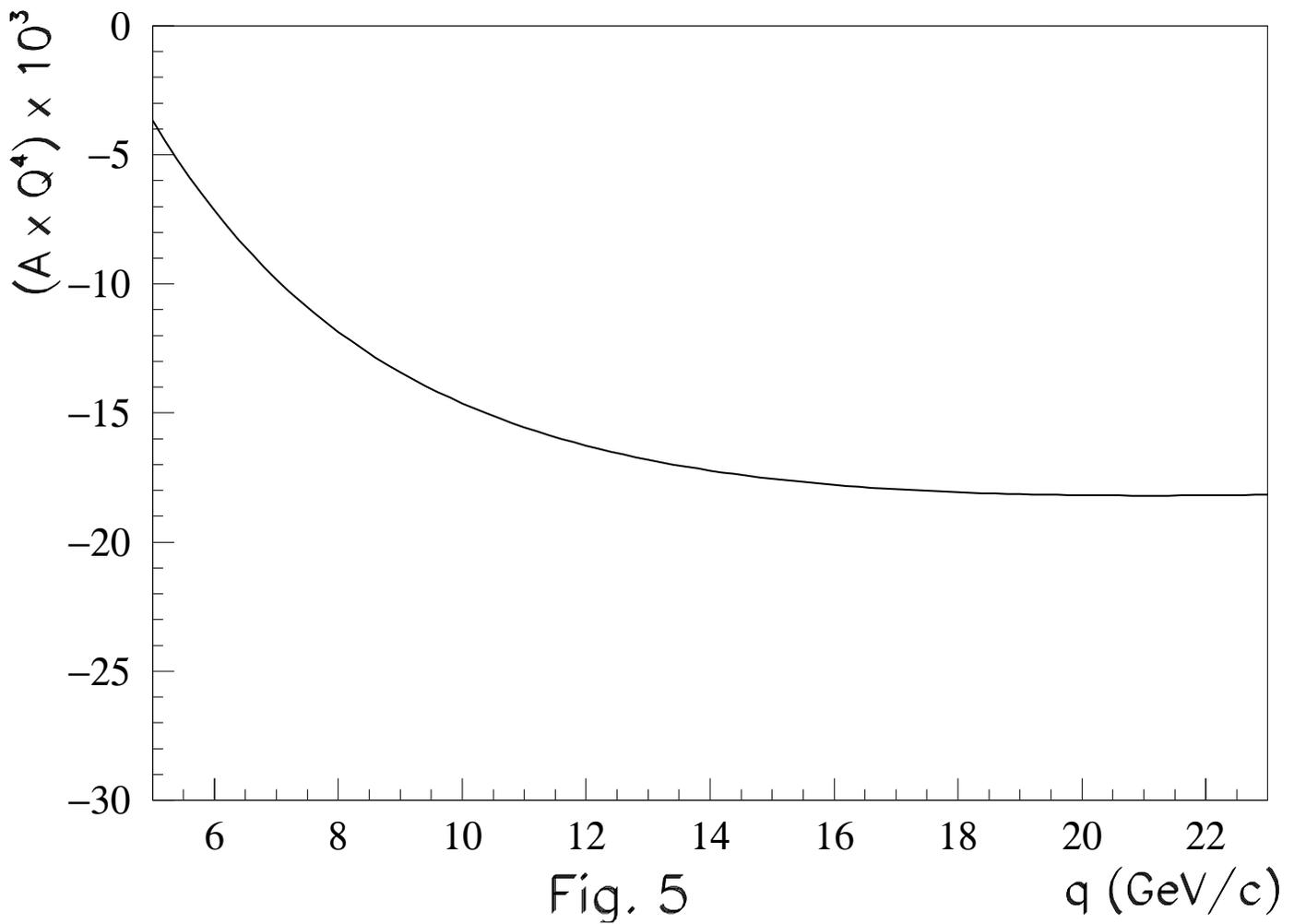
\begin{figure}
\caption{The product $A * Q^4$, multiplied by $10^3$, as a function of 
the momentum transfer $q$ in the range $5 \leq q \leq 23$ GeV/$c$ for 
the 
$^{12}\protect{\text{C}}(\vec e,e'p)^{11}\protect{\text{B}}_{s 
\scriptstyle{1 \over 2}}$ reaction in the same kinematical conditions 
as for the solid line in Fig.\ \protect{\ref{fig2}}.}
\label{fig5}
\end{figure}


\begin{thebibliography}{10}

\bibitem{domb71}
N. Dombey,  in {\em Hadronic Interactions of Electrons and Photons}, 
edited by  J. Cumming and H. Osborn (Academic Press, London, 1971), 
p.\ 17.

\bibitem{bgp85}
S. Boffi, C. Giusti, and F.~D. Pacati, Nucl. Phys. {\bf A435},  697  
(1985).

\bibitem{donn85}
T.~W. Donnelly, Prog. Part. Nucl. Phys. {\bf 13},  183  (1985).

\bibitem{bgprbook}
S. Boffi, C. Giusti, F.~D. Pacati, and M. Radici, {\em Electromagnetic 
Response  of Atomic Nuclei}, Vol.~20 of {\em Oxford Studies in Nuclear 
Physics} (Oxford  University Press, Oxford, 1996).

\bibitem{frumou}
S. Frullani and J. Mougey, Adv. Nucl. Phys. {\bf 13},  1  (1984).

\bibitem{simo71}
M. Simonius,  in {\em Polarization Phenomena in Nuclear Reactions}, 
edited by  H.~H. Barschall and W. Haeberli (University of Wisconsin 
Press, Madison,  1971), p.\ 401.

\bibitem{donn84}
T.~W. Donnelly,  in {\em Perspectives in Nuclear Physics at 
Intermediate  Energies}, edited by S. Boffi, C.~C. degli Atti, and 
M.~M. Giannini (World  Scientific, Singapore, 1984), p.\ 244.

\bibitem{oops94}
J. Mandeville {\it et~al.}, Phys. Rev. Lett. {\bf 72},  3325  (1994).

\bibitem{papa89}
C.~N. Papanicolas,  in {\em Excited Baryons 1988}, edited by G. Adams, 
N.~C.  Mukhopadhyay, and P. Stoler (World Scientific, Singapore, 1989), 
p.\ 235.

\bibitem{donrask86}
T.~W. Donnelly and A.~S. Raskin, Ann. Phys. (NY) {\bf 169},  247  
(1986).

\bibitem{rob90}
R.~G. Roberts, {\em The Structure of the Proton. Deep Inelastic 
Scattering}  (Cambridge University Press, Cambridge, 1990).

\bibitem{elfeproj}
 in {\em Conference Proceedings of the Italian Physical Society: the 
 ELFE  Project, an Electron Laboratory for Europe}, edited by J. 
 Arvieux and E.~D.  Sanctis (Editrice Compositori, Bologna, 1993).

\bibitem{cebaf91006}
A. Saha~(spokesperson), CEBAF Proposal No. E-91-006  (1991).

\bibitem{cebaf91007}
R. Milner~(spokesperson), CEBAF Proposal No. E-91-007  (1991).

\bibitem{cebaf91013}
D.~F. Geesaman~(spokesperson), CEBAF Proposal No. E-91-013  (1991).

\bibitem{framilstri94}
L.~L. Frankfurt, G.~A. Miller, and M.~I. Strikman, Annu. Rev. Nucl. 
Part. Sci.  {\bf 45},  501  (1994).

\bibitem{niko94}
N.~N. Nikolaev, Int. J. Mod. Phys. {\bf E3},  1  (1994).

\bibitem{bbkha94}
A. Bianconi, S. Boffi, and D.~E. Kharzeev, Yadernaya Fizika {\bf 57},  
1732  (1994).

\bibitem{glau59}
R.~J. Glauber,  in {\em Lectures in Theoretical Physics}, edited by W. 
Brittain  and L.~G. Dunham (Interscience Publ., New York, 1959), 
Vol.~1.

\bibitem{pprev78}
G.~D. Alkhazov, S.~I. Belostotsky, and A.~A. Vorobyev, Phys. Rep. 
{\bf 42},  89   (1978).

\bibitem{br96a}
A. Bianconi and M. Radici, Phys. Rev. {\bf C53},  R563  (1996).

\bibitem{bccgp82}
S. Boffi {\it et~al.}, Nucl. Phys. {\bf A379},  509  (1982).

\bibitem{fw50}
L.~L. Foldy and S.~A. Wouthuysen, Phys. Rev. {\bf 78},  29  (1950).

\bibitem{gp80}
C. Giusti and F.~D. Pacati, Nucl. Phys. {\bf A336},  427  (1980).

\bibitem{bosted95}
P.~E. Bosted, Phys. Rev. {\bf C51},  409  (1995).

\bibitem{br95}
A. Bianconi and M. Radici, Phys. Lett. {\bf B363},  24  (1995).

\bibitem{br96b}
A. Bianconi and M. Radici, Phys. Rev. {\bf C54},  3117  (1996).

\bibitem{ne18}
N.~C.~R. Makins {\it et al.}~(the NE18~Collaboration), Phys. Rev. Lett. 
{\bf  72},  1986  (1994).

\bibitem{comf80}
J.~R. Comfort and B.~C. Karp, Phys. Rev. {\bf C21},  2162  (1980).

\bibitem{brolep81}
S.~J. Brodsky and G.~P. Lepage, Phys. Scri. {\bf 23},  945  (1981).

\bibitem{bb95}
A. Bianconi and S. Boffi, Phys. Lett. {\bf B348},  7  (1995).

\bibitem{davy}
A.~S. Davydov, {\em Kvantovaja Mechanika} (Nauka, Moscow, 1981).

\end{thebibliography}
\end{document}